\documentclass[showkeys,prd,onecolumn,preprintnumbers,floats,floatfix,apsrev,amsmath,amsfonts,nofootinbib,superscriptaddress]{revtex4-2}
\usepackage{color}
\usepackage{graphicx}
\usepackage{dcolumn}
\usepackage{natbib}
\usepackage{hyperref}
\usepackage{amsthm}%
\usepackage{hypernat}
\usepackage{natbib}
\usepackage{soul}
\usepackage{amssymb}
\usepackage{orcidlink}

\newtheorem{thm}{Theorem} 
\newcommand{\dd}{{\rm d}}

\newcommand{\p}{\partial}

\newcommand{\be}{\begin{equation}}
\newcommand{\en}{\end{equation}}

\newcommand{\paren}[1]{\left({#1}\right)}

\begin{document}
\title{Weak Cosmic Censorship in Einstein-Born-Infeld Black Hole: A Parameter-Space Analysis}

\author{M.~L.~Pe\~{n}afiel\,\orcidlink{0000-0002-7700-1787}}
\email{miguelpenafiel@upb.edu}
\affiliation{FIA - Facultad de Ingeniería y Arquitectura, Universidad Privada Boliviana, Camino Achocalla Km 3.5, La Paz, Bolivia.}
\affiliation{Carrera de Ciencias Físicas y Energías Alternativas, Universidad Pública de El Alto, Zona Villa Esperanza Av. Sucre B S/N, El Alto, Bolivia.}

\author{F.~T.~Falciano\,\orcidlink{0000-0003-2263-1252}}
\email{ftovar@cbpf.br}
\affiliation{CBPF -- Brazilian Center for Research in Physics, R.~Xavier Sigaud 150, 22290-180, Rio de Janeiro, RJ, Brazil}
\affiliation{PPGCosmo/CCE -- Federal University of Esp\'irito Santo, 29075-910, Vit\'oria, ES, Brazil}

\author{J.~C.~Fabris\,\orcidlink{0000-0001-8880-107X}}
\email{julio.fabris@cosmo-ufes.org}
\affiliation{PPGCosmo/CCE -- Federal University of Esp\'irito Santo, 29075-910, Vit\'oria, ES, Brazil}
\affiliation{National Research Nuclear University MEPhI, Kashirskoe sh. 31, 115409 Moscow, Russia}

\begin{abstract}
Einstein-Born-Infeld is a charged black hole solution of general relativity with Born-Infeld nonlinear electrodynamics, which has interesting physical features. It violates Bekenstein's entropy upper bound, and it can overcharge with respect to the Reissner-Nordstr\"om black hole. It presents an optimum situation to test the cosmic censorship conjecture, which is another important conjecture in black hole physics. In principle, this analysis requires the calculation of the self-energy of a test particle in the vicinity of the black hole, which, in general, is not feasible because of the electromagnetic nonlinearities. In this work, we propose an alternative approach that allows us to explore the whole parameter space of the Einstein-Born-Infeld black hole and test the validity of the cosmic censorship conjecture.
\end{abstract}


\maketitle

\section{Introduction}
The cosmic censorship conjecture, first proposed by Penrose~\cite{Penrose1969}, plays a central role in black hole mechanics. The conjecture asserts that the inevitable formation of singularities in general relativity must be accompanied by an event horizon, which prevents the loss of predictability of the theory. The stability of black hole solutions, the universality of the no-hair theorem, and the thermodynamic analogy all implicitly rely on the assumption that singularities are not exposed. Without the validity of cosmic censorship, these cornerstones of black hole physics could require substantial revision, with eventually far-reaching consequences for both classical and quantum gravity.

The conjecture is also deeply connected to the generalized second law~\cite{Bekenstein1972,Wald1984,Bekenstein1974}, namely that the sum of the entropy of the black hole and matter fields can never decrease. The area increase theorem, for example, presupposes the regularity of horizons and the causal disconnection of singularities from the asymptotic region. The breakdown of cosmic censorship undermines the analogy that has proven to be so fruitful in the development of quantum gravity and quantum field theory in curved spacetime such as in Hawking radiation and holography.

A common strategy for testing the WCCC consists in examining whether an extremal or near-extremal black hole can be driven beyond extremality through the absorption of matter. Since the pioneering work of Wald \cite{Wald1974}, numerous gedanken experiments have been developed to examine the WCCC validity in a variety of gravitational theories and black hole solutions \cite{Hod1999,Hod2000,Wald1974,Wald1999,Hod2008,Hubeny1999,Bekenstein1994,deFelice2001,Hod2002,Matsas:2007bj,Richartz:2011vf,Lu:2025efo,Zeng:2019jta,Wang:2019jzz,Hu:2019rpw}. 

Among the many extensions of Einstein-Maxwell theory, Born-Infeld electrodynamics is a widely studied theory originally introduced in order to regularize the self-energy and electric field of a point particle \cite{Born1933,Born410,Born425}. It stands as a distinctive example due to its unique features: it is the only theory with absence of birefringence that respects the adequate weak field limit \cite{Boillat1970,Plebanski1970,Russo2023}, it arises in the low-energy limit of string theory \cite{Fradkin1985} and is a completely regular theory.

When coupled to General Relativity, the Einstein-Born-Infeld black hole solution can display different horizon and singularity configurations, offering the possibility of having black hole solutions which are overcharged with respect to the extremal Reissner-Nordstr\"{o}m solution \cite{Fernando2006,Chemissany2008}. Furthermore, it has been shown that generalized versions of Bekenstein’s entropy bound (c.f. \cite{Bekenstein1981,Zaslavskii1992,Zaslavskii1996,Hod1999a,Hod1999b,Bekenstein1999,Schiffer1989,Casini2008,Blanco2013,Myers2012,Bekenstein1987,Schiffer1990,Carullo:2021yxh,Dain:2015ira,Penafiel2017,Diaz:2024wbh,Longo:2024gly,Buoninfante:2020guu}) may be violated in the presence of Born-Infeld nonlinearities \cite{Falciano2019,Falciano2021}. These results suggest that nonlinear electromagnetic effects can substantially alter the relationship between charge, energy, entropy, and horizon formation that characterizes Einstein-Maxwell black holes.

The fact that Einstein-Born-Infeld black holes exhibit departures from Einstein-Maxwell theory regarding the extremality conditions for overcharging the black hole and violate the entropy bounds raises the question whether other results stemming from black hole physics may be affected. In particular, one may ask whether the same nonlinearity effects giving rise to modifications of the entropy bounds and allowing a black hole to have a larger charge-to-mass ratio than in the Reissner-Nordstr\"{o}m case may also affect the mechanisms responsible for preserving the event horizon once a near-extremal black hole absorbs a charged particle, i.e., if WCCC is preserved for these black holes.

Several approaches may be employed to answer this question. In particular, the validity of the WCCC for Einstein-Born-Infeld black holes been addressed by He and Jiang in \cite{He2019} using the Sorce-Wald formulation of gedanken experiments \cite{Sorce2017}, where first and second order inequalities are obtained for WCCC preservation of these black holes, as well as extensions of the WCCC validity within Born-Infeld modified gravity \cite{Jana2018}. In the present work, we study the validity of WCCC for Einstein-Born-Infeld black holes by considering charged particle absorption by a near-extremal black hole. In order to inspect this situation, we derive the electrostatic potential of a point particle in this theory and reformulate the WCCC violation condition into an inequality relating the direction of the physical displacement within the phase space and the slope of the curve defining extremal solutions. Thus, we provide a geometrical criterion for WCCC violation within the phase space for black hole configurations that can be extended to other theories of nonlinear electrodynamics.

This work is organized as follows. In Sec. \ref{sec:NLEDcst} we review the main properties of Born-Infeld nonlinear electrodynamics in curved spacetime and summarize the relevant features of the electric field due to a point particle in such a curved spacetime. In Sec. \ref{sec:EBIBH} discuss the black hole solution for Born-Infeld electrodynamics coupled to General Relativity, as well as introduce the phase structure determining the horizon and singularity behavior for this solution. In Sec. \ref{sec:WCCC} we analyze the validity of the WCCC by considering how the absorption of charged matter induces a displacement in the parameter space. Finally, in Sec. \ref{sec:concl} we discuss the implications of our results and present further research possibilities for the method presented here.

\section{Born-Infeld Electrodynamics in Curved Spacetime} \label{sec:NLEDcst}

Electromagnetism is a vector gauge theory for the $U(1)$ symmetry group. As a consequence, the Faraday tensor is given in terms of the vector potential $F_{\mu\nu}=\partial_{\mu}A_{\nu}-\partial_{\nu}A_{\mu}$, and the dual of the Faraday tensor is $\tilde{F}^{\mu\nu}=\frac12\eta^{\mu\nu\alpha\beta}F_{\alpha\beta}$ where $\eta^{\mu\nu\alpha\beta}$ is the totally antisymmetric Levi-Civita tensor. The covariant electric and magnetic fields are projections of the Faraday tensor and its dual along the normalized observer's worldline. An observer $\mathbf{v}$ measures an electric field $\mathbf{E}$ and magnetic field $\mathbf{B}$ given by
\begin{align}\label{EBF}
E^\mu=F^{\mu}{}_{\alpha}v^{\alpha}/c \quad , \quad B^{\mu}=\tilde{F}^{\mu}{}_{ \alpha}v^{\alpha}/c
\quad .
\end{align}

These are pure spatial vectors in the sense that they are orthogonal to the observer worldline, i.e. $\mathbf{E}\cdot \mathbf{v}=\mathbf{B}\cdot \mathbf{v}=0$. There are only two linearly independent Lorentz invariants constructed with the Faraday tensor and its dual, namely $F\equiv\frac12F^{\mu\nu}F_{\mu\nu}=\mathbf{E}\cdot \mathbf{E}
-\mathbf{B}\cdot \mathbf{B}$ and $G\equiv \frac12\tilde{F}^{\mu\nu}F_{\mu\nu}=2\mathbf{E}\cdot \mathbf{B}$. A generic nonlinear electrodynamic can be specified by its Lagrangian density $\mathcal{L}\left(F,G\right)$. In particular, the Born-Infeld theory has the following action
\begin{equation} \label{eq:LBI}
S=\frac{1}{4\pi}\int \dd^4 x\sqrt{-g} \ \beta^2\left(1-\sqrt{
1+\frac{F}{\beta^2}-\frac{G^2}{4\beta^4}
}\right)\quad ,
\end{equation}
where $\beta $ is the parameter that specifies the maximum value of the field strength. Nonlinear electrodynamics has been studied in several physical situations~\cite{Breton:2004qa,Wei:2021vdx,Bronnikov:2000vy,Rasheed:1997ns,Diaz:2024wbh}. Born-Infeld is a nonlinear electrodynamics that approaches Maxwell theory in the weak field limit ($|\mathbf{E}|$ and $|\mathbf{B}|\ll\beta$). It satisfies the experimental constraints depending on the value of the field strength parameter $\beta$. This theory was originally proposed by Born and Infeld ~\cite{Born1933,Born410,Born425} as a means to avoid the classical divergences in Maxwell theory. It is unique in the sense that it is the only nonlinear electrodynamics without birefringence \cite{Boillat1970,Plebanski1970}, and a perturbative expansion reproduces the Euler-Heisenberg electrodynamic Lagrangian, which is an effective theory for quantum electrodynamics when vacuum polarization effects are taken into account~\cite{Euler1936}. The dynamics reads
\begin{align}\label{fNLED}
\nabla_{\mu}E^{\mu\nu}&=-4\pi j^{\nu} \ ,\quad\\
E^{\mu\nu}&\equiv -2\left(\mathcal{L}_FF^{\mu\nu}+\mathcal{L}_G\tilde{F}^{\mu\nu}\right) \ ,\notag 
\end{align}
where $E^{\mu\nu}$ is the excitation tensor and $\mathcal{L}_x$ represents the derivative of $\mathcal{L}$ with respect to $x$. The decomposition of the excitation tensor defines the electric displacement $\mathbf{D}$ and the magnetic field $\mathbf{H}$ as a projection along the observer's worldline. Similarly to the decomposition \eqref{EBF}, $D^\mu=E^{\mu}{}_{\alpha} v^{\alpha}/c$ and $H^{\mu}=\tilde{E}^{\mu}{}_{ \alpha}v^{\alpha}/c$, hence we have the following constitutive relations.
\begin{align}\label{DHE}
\mathbf{D}&=-2\mathcal{L}_F\mathbf{E} -2\mathcal{L}_G\mathbf{B}
\\
\mathbf{H}&=-2\mathcal{L}_F\mathbf{B} +2\mathcal{L}_G\mathbf{E}
\end{align}

Generically, the fields $\mathbf{E}$ and $\mathbf{B}$ do not share the same symmetries and physical properties of $\mathbf{D}$ and $\mathbf{H}$. In particular, an electrostatic configuration implies $\mathbf{E}=-\boldsymbol{\nabla}\phi(x)$ but in general $\boldsymbol{\nabla}\times \mathbf{D}\neq 0$. Notwithstanding, it has been shown~\cite{Falciano2019,Falciano2021} that the displacement vector of an arbitrary nonlinear electrodynamic satisfies the theorem.


\begin{thm}\label{thmNLEDSSBH}
The electrostatic potential $\phi(x)$ produced by a charged particle that satisfies a nonlinear electrodynamic theory $\mathcal{L}(F,G)$ in a static spherically symmetric spacetime is entirely specified by the electrostatic potential $\psi(x)$ with the same boundary conditions but satisfying Maxwell electromagnetism in the same background. The displacement vector is curl-free and is given by $\mathbf{D}=-\boldsymbol{\nabla}\psi(x)$.
\end{thm}

\begin{figure}[t]
\centering
\includegraphics[width=0.5\textwidth]{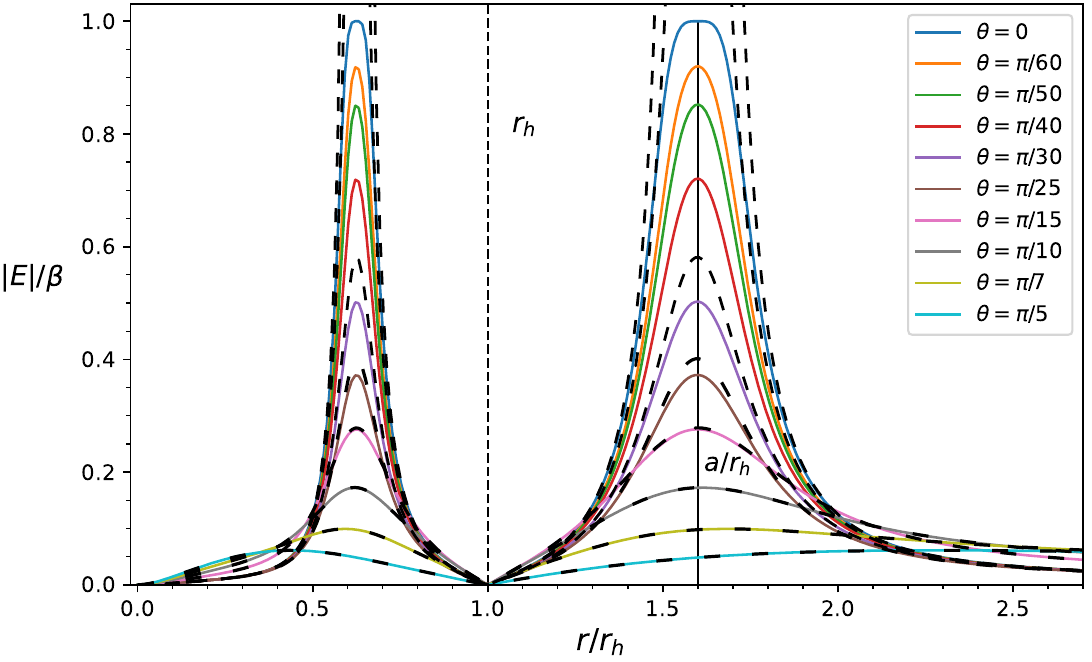}
\caption{Plot of the  Born-Infeld (solid) and Maxwell (dashed) electric field of a point-like particle outside a Schwarzschild Black Hole. The particle is located at $\left(\theta=0,a=1.6 r_h\right)$ and $\eta =1/32\varepsilon^2$. The Born-Infeld field is finite everywhere and coincides with the Maxwell field away from the particle. The field's peak decreases and shifts away from the horizon with increasing values of $\theta$. }
	\label{fig:BI_Max_EField}
\end{figure}

Electrostatic means $\mathbf{B}=0$ and $\partial_0\mathbf{E}=0$. Thus, $\mathbf{E}=\left(0, -\boldsymbol{\nabla}\phi \right)$ in the reference frame where the particle is at rest, and the electric displacement becomes $\mathbf{D}=-2 \mathcal{L}_F(E)\, \mathbf{E}$. Now, \eqref{fNLED} simplifies to
\begin{align}\label{NLEDisotropic}
	\mathbf{\nabla}\cdot \mathbf{D} =-4\pi \rho \quad ,
\end{align}
where the charge density is $\rho=\mathbf{j}\cdot\mathbf{v}/c^2$. In a static spherically symmetric spacetime, the velocity of the particle at rest is $\mathbf{v}=(c/\sqrt{g_{00}},0,0,0)$, and hence $\mathbf{\nabla} \cdot \mathbf{v} = \mathbf{\partial} \cdot \mathbf{v}=0$. Note that \eqref{NLEDisotropic} is formally identical to Maxwell's theory with solution $\mathbf{D}=-\boldsymbol{\nabla}\psi$, and setting 
\begin{align} \label{def:auxfunc}
	\psi ( x)&= -2 \int  \ \mathcal{L}_F(\boldsymbol{\nabla}\phi) \boldsymbol{\nabla}\phi \cdot \mathbf{\dd l}
\end{align}
we automatically satisfy $\mathbf{D}=2 \mathcal{L}_F(\boldsymbol{\nabla}\phi)\, \boldsymbol{\nabla}\phi=-\boldsymbol{\nabla}\psi$. As a consistent check, we can verify the Bianchi identity, which in the electrostatic case reduces to $\boldsymbol{\nabla}\times\mathbf{E}=0$. Inverting the constitutive relation, we find $\mathbf{E}=-f\left(|\boldsymbol{\nabla}\psi|\right)\boldsymbol{\nabla}\psi$, where $f\left(|\boldsymbol{\nabla}\psi|\right)$ depends on the specific nonlinear electrodynamic. Notwithstanding, what is important is that
\begin{equation}
\boldsymbol{\nabla}f =\frac{\partial f}{\partial \ln D} \ (\boldsymbol{D} \cdot \boldsymbol{\nabla}) \boldsymbol{D} 
\end{equation}
which is collinear to $\boldsymbol{D}$. Therefore, $\boldsymbol{\nabla}\times\mathbf{E}=f\boldsymbol{\nabla}\times\mathbf{D}+\boldsymbol{\nabla}f\times\mathbf{D}=0$, which proves the theorem. This result has interesting consequences for charged particles in the vicinity of a black hole. The constitutive relations for electrostatic fields in Born-Infeld read
\begin{align}\label{constEqsESt}
	\mathbf{D}= \frac{\mathbf{E}}{\sqrt{1- |\mathbf{E}|^2 \beta^{-2} }} \quad , \quad 
	\mathbf{E}= \frac{\mathbf{D}}{\sqrt{1+ |\mathbf{D}|^2 \beta^{-2} }} \quad . 
\end{align}
Theorem~1 guarantees that in the vicinity of a static spherically symmetric black hole $\mathbf{E}(r,\theta)=-\boldsymbol{\nabla}\phi
(r,\theta)$ with 
\be\label{constitBI2}
\boldsymbol{\nabla}\phi(r,\theta)=\frac{\boldsymbol{\nabla}\psi(r,\theta)}{\sqrt{1+|\boldsymbol{\nabla}\psi(r,\theta)|^2\beta^{-2}}}\ , 
\en
where $\psi(r,\theta)$ is Copson's solution that describes a Maxwell electrostatic potential in the same spacetime (see \cite{Falciano2019,Falciano2021}). Although $|\boldsymbol{\nabla}\psi|$ diverges, the electric field is finite everywhere, even over the particle. Figure~\eqref{fig:BI_Max_EField} shows the electric field of a point-like source for Maxwell theory (dashed curves) and Born-Infeld theories (solid curves) in the vicinity of a Schwarzschild black hole. As expected, far from the particle both solutions coincide. Note also that both fields vanish at the horizon, which is consistent with interpreting the horizon as a conducting surface~\cite{Price1986,MacDonald1985}. As the particle approaches the horizon, the Born-Infeld regions, defined as $|\nabla\psi|\geq \beta$, are squeezed and piled but never touch the horizon, which is compatible with the vanishing of the electric field at the horizon. The conducting behavior of the horizon makes the horizon polarization invariant with respect to the nonlinear electromagnetic theory, as long as the theory has the correct Maxwell weak field limit.\\

\section{Einstein-Born-Infeld Black Hole} \label{sec:EBIBH}

The Einstein-Born-Infeld black hole is a solution to the coupled system of general relativity and Born-Infeld electrodynamics that describes a static spherically symmetric spacetime~\cite{Demianski1986,Garcia1984}. In standard coordinates, the line element is
\[
\dd s^2=g_{00}(r)c^2\dd t^2-g_{00}(r)^{-1}\dd r^2-r^2\left(\dd\theta^2+\sin^2 \theta \dd\varphi^2 \right)\ .
\]

It has three parameters: the mass $M$ and the charge $q$ of the black hole, and the Born-Infeld parameter $\beta$. We can recast them as characteristic lengths by defining the following quantities.
\begin{equation} \label{defrs}
r_s\equiv \frac{2GM}{c^2}\ , \quad r_q\equiv \frac{\sqrt{G}q}{c^2}\ , \quad r_\beta\equiv \frac{c^2}{\sqrt{G}\beta}\ ,
\end{equation}
where $c$ and $G$ are, respectively, the speed of light and Newton's gravitational constant. The Einstein-Born-Infeld black hole has four distinct behaviors depending on the smallness of two dimensionless parameters.
\begin{equation}
\eta \equiv {r_\beta}/{r_q} \qquad , \qquad \varepsilon \equiv {r_q}/{r_s} \quad .
\end{equation}
The parameter $\eta$ measures the importance of electromagnetic nonlinearities (deviations from Maxwell's theory). For $\eta\ll1$, the nonlinearities are negligible and the solution is close to Reissner-Nordstr\"om, hence $\eta\ll1$ corresponds to the superficial Born-Infeld regime. In the opposite limit, namely $\eta > 1$, the nonlinearities are predominant, and we have the deep Born-Infeld regime. The parameter $\varepsilon$ controls how much the electromagnetic energy contributes to the spacetime curvature, that is, the charge-to-mass ratio of the object. This parameter ranges from $0\le\varepsilon\le\varepsilon_{ex}$. For $\varepsilon\ll1$ the black hole mass is the main source for gravity while for $\varepsilon\sim\varepsilon_{ex}$, the electromagnetic contribution to the black hole energy is maximum, and we have an extreme black hole. For all cases, the metric reads
\begin{align*}
g_{00}(r)&=1-\frac{r_s}{r}+\frac{r_q^2}{r^2} f\left(\frac{r}{ \varepsilon \sqrt{\eta} \, r_s }\right)\ , \\ 
f(x)&\equiv\frac{2}{3}x^4\left(1-\sqrt{1+\frac{1}{x^4}}\right)+\frac{4}{3}\ _2F_1\left[\frac{1}{2},\frac{1}{4},\frac{5}{4},-\frac{1}{x^4}\right]\ , 
\end{align*}
with the black hole's electrostatic potential 
\be
\phi_q(r)=\frac{q}{r}\ _2F_1\left[\frac{1}{2},\frac{1}{4};\frac{5}{4};-\left( \frac{ \varepsilon \sqrt{\eta}\,   r_s}{r}\right)^4\right]\ , \label{EBIeletrostaticpot}
\en
where $_2F_1\left[a,b;c;z\right]$ stands for the Gaussian hypergeometric function. The exterior horizon is a solution to $g_{00}(r_h)=0$ and has an implicit expression given by
\[
r_{h}=\frac{r_s}{2}\left[1+\sqrt{1-4\varepsilon^2 f\left(\frac{r_h}{ \varepsilon \sqrt{\eta}\,r_s}\right)} \ \right]\ .
\]


\begin{figure}[t]
\centering
\includegraphics[width=0.5\textwidth]{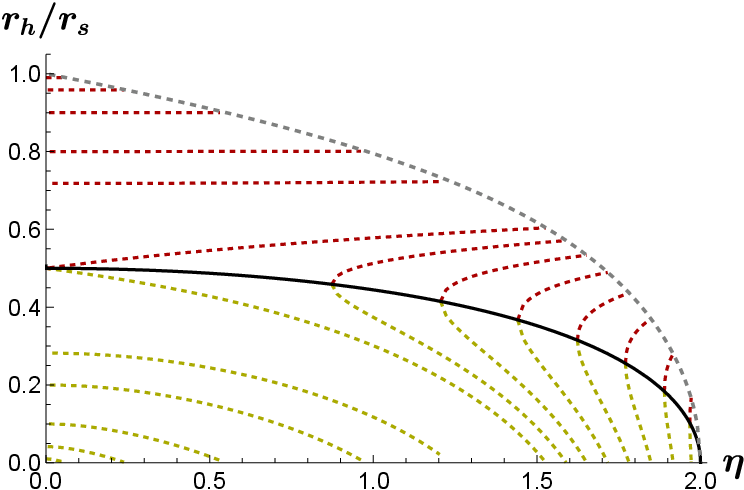}
\caption{We plot the inner (yellow) and outer (red) horizon radii, for fixed values of $\varepsilon$, as a function of $\eta$ for the RN phase. The black curve represents the extreme Einstein-Born-Infeld black hole as a function of $\eta$. Above the gray curve, we have a naked singularity. From left to right: $\varepsilon=\paren{0.1,0.2,0.3,0.4,0.5,0.51,0.52,0.53,0.54,0.55,0.56,0.568}$.}\label{fig:rhrs}
\end{figure}

There is no analytic solution for $r_{h}=r_h(r_s,\varepsilon,\eta)$ but it can be found numerically (see figure~\eqref{fig:rhrs}). For fixed values of the parameters, we can expand the solution close to the origin. We find
\begin{align*}
\lim_{r\to0}g_{00}&=1-\frac{2}{\eta}-\frac{r_s}{r}\left(1-\frac{\Gamma\left(1/4\right)^2}{3\sqrt{\pi}}\frac{\varepsilon}{\sqrt{\eta}}\right)+\mathcal{O}\left(\frac{r}{r_q \eta}\right)^2 \ . 
\end{align*}

Depending on the values of $\varepsilon$ and $\eta$, the singularity can be spacelike or timelike. An interesting situation arises for
\be\label{eq:critical}
\varepsilon=\varepsilon_\ast \sqrt{\frac{\eta}{2}}
\quad , \quad \varepsilon_\ast\equiv 
\frac{3\sqrt{2\pi}}{\Gamma\paren{1/4}^2}\approx 0.57207
\ ,
\en
in which case, assuming $\eta\ne0$, there is no singularity at the origin. Eq. \eqref{eq:critical} defines a critical line in parameter space. For $\varepsilon<\varepsilon_\ast \sqrt{\eta/2}$, the singularity is spacelike similar to the Schwarzschild and the black hole has only one horizon. For $\varepsilon>\varepsilon_\ast\sqrt{\eta/2}$, we have the Reissner-Nordstr\"om behavior, where the singularity is timelike and the black hole can have two, one, or no horizons, respectively, for the cases of non-extreme, extreme, and naked singularity. In the domain $\eta\in[0,2)$, the critical line \eqref{eq:critical} divides the Reissner-Nordstr\"om-like phase and the Schwarzschild-like phase, while for $\eta>2$ this line divides the Schwarzschild phase and the naked singularity configuration. The condition for finding the extreme horizon radius is $\p_r g_{00}=g_{00}=0$ which gives
\begin{align} 
r_{ex}&=
r_s\, \varepsilon \sqrt{1-\frac{\eta^2}{4}} \quad  ,
\quad  f_{ex}\paren{\eta}=f\paren{\sqrt{\frac{1}{\eta}-\frac{\eta}{4}}}
\ ,
\notag\\
\varepsilon_{ex}&=\frac{\sqrt{1-\eta^2/4}}{1-\eta^2/4+f_{ex}\paren{\eta}}
\ . 
\label{eq:rex1}
\end{align}

\begin{figure}[t]
\centering
\includegraphics[width=0.5\textwidth]{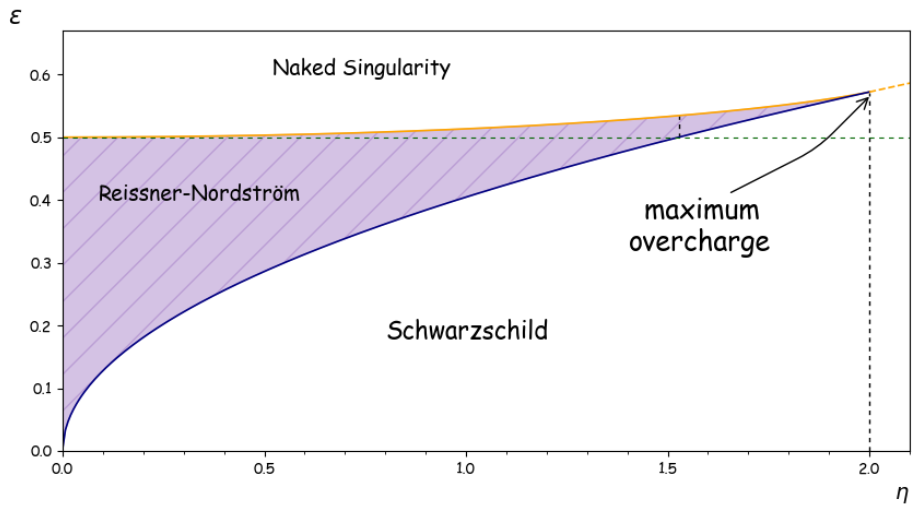}
\caption{We display the phase diagram for the Einstein-Born-Infeld black hole. The Reissner-Nordstr\"om phase is delimited by the extremality condition given by Eq. \eqref{eq:rex1} 
(yellow line) and the existence condition for the Reissner-Nordstr\"om phase (blue line). For $\eta>2$ (yellow dotted line), only black holes within the Schwarzschild phase are allowed.}\label{fig:phasediag}
\end{figure}

In the limit $\eta\to0$, Einstein-Born-Infeld black hole is indistinguishable from Reissner-Nordstr\"om black hole and the extreme black hole has $\varepsilon_{ex}=1/2$. As $\eta$ increases, the charge-to-mass ratio increases and the extreme Einstein-Born-Infeld black hole becomes overcharged relative to the extreme Reissner-Nordstr\"om case. The maximum of overcharge happens for $\eta\to2$, hence the extreme Einstein-Born-Infeld black hole has
\begin{equation}\label{epsrange}
0.5\le\varepsilon_{ex}<\varepsilon_\ast \ .
\end{equation}

Figure~\eqref{fig:phasediag} shows the three possible configurations for the Einstein-Born-Infeld black hole. The usual Reissner-Nordstr\"om black hole lies on the line $\eta=0$. As $\eta$ increases, the different phases emerge. The extreme line gives the upper limit of overcharging for the Reissner-Nordstr\"om phase. As $\eta\to2$, both the extreme line and the lower Reissner-Nordstr\"om phase line coincide. This point is a type of \emph{triple point}, where the three Einstein-Born-Infeld black hole phases coexist. For $\eta>2$, the electrodynamics nonlinearities dominate and suppress the effect of the electric charges, that is, there is only a Schwarzschild-like black hole.\\

\section{Weak Cosmic Censorship Conjecture} \label{sec:WCCC}

The fact that Born-Infeld nonlinear electrodynamics violates Bekenstein's inequalities~\cite{Falciano2019,Falciano2021} prompts us to check if it also violates another important conjecture in black hole physics, namely the cosmic censorship conjecture. This conjecture claims that no physical process can yield a naked singularity~\cite{Penrose1969}. A weak version of the conjecture states that all singularities must be dressed by a horizon. A common way to test it is to let a near-extreme black hole absorb some matter field and look for possible ways to turn the black hole into a naked singularity~\cite{Wald1974,Wald1999,Hod2008,Bekenstein1994,Hod1999,Hubeny1999,Hod2000,deFelice2001,Hod2002,Matsas:2007bj,Richartz:2011vf,Lu:2025efo,Zeng:2019jta,Wang:2019jzz,Hu:2019rpw}. 

The Einstein-Born-Infeld black hole can be overcharged relative to Reissner-Nordstr\"om black hole; therefore, a near-extreme Einstein-Born-Infeld black hole is a good candidate to violate the weak cosmic censorship conjecture (see Figure~\ref{fig:phasediag}). In particular, nonlinear effects are stronger for large values of $\eta \lesssim 2$, where we expect larger deviations from the conventional Reissner-Nordstr\"om black hole. 

By definition, a black hole is extreme if $g_{00}=\p_r g_{00}=0$, with solution \eqref{eq:rex1}. In the limit $\eta\rightarrow 0$, the horizon tends to $r_{ex}\rightarrow  r_s/2$ and for the region outside the black hole the argument of the function $f_{ex}(\eta)$ is much larger than one; hence, as expected, everything goes as an extreme Reissner-Nordstr\"om black hole. However, as $\eta$ increases, not only the argument of the function $f(x)$, but also the size of the horizon $r_{ex}$ changes significantly.

In order to test the cosmic censorship conjecture, the idea is to lower a small charged body close to the horizon and drop it into the black hole. A horizon exists only for $\varepsilon \leq \varepsilon_{ex}$. After absorbing a particle of energy $\mathcal{E}$ and charge $e$, the black hole mass and charge change to $M\to M+\mathcal{E}/c^2$ and $q\to q+e$. 

The energy $\mathcal{E}$ depends on the electrostatic potential of the test charged body approaching the black hole. Calculating it is a nontrivial task due to the nonlinearity of the constitutive relations for the Born-Infeld electromagnetic fields \eqref{constEqsESt}. The issue is that Born-Infeld nonlinearities forbid us to find a closed form for the Green function. In other words, we do not have an analytical solution for the electrostatic potential. In the slowly lowering regime the test body's energy reads
\begin{equation}\label{Eptest}
\mathcal{E}= mc^2	\sqrt{g_{00}(a)}	+e\, \phi_q(a)+\frac{e}{2} \phi_e(a)
\ ,
\end{equation}
where the first two terms are respectively the rest mass energy and the electrostatic energy of the test body with charge $e$ in the electrostatic potential $\phi_q(r,\theta)$ produced by the black hole. The last term $e\,\phi_e(r,\theta)$ is associated with the self-energy of the test body, which is sensitive to electromagnetic nonlinearities and does not have an analytical solution. In certain situations, one can find valuable approximate solutions~\cite{Falciano2021}; however, they are valid in a limited region of the Einstein-Born-Infeld parameter space. In order to analyze the validity of the cosmic censorship conjecture for the whole parameter space, and circumvent this limitation, we shall use a different approach that does not require the calculation of the particle's electrostatic self-energy. 

The change in the mass and charge of the black hole causes a simultaneous change in the parameters $\varepsilon$ and $\eta$ given by
\begin{align}
\Delta \varepsilon&= \frac{e}{q}\varepsilon\left(\frac{1-\varepsilon\,V_0}{1+\varepsilon\,V_0\, e/q}\right) \quad , \quad  \Delta \eta=-\eta\left(\frac{e/q}{1+e/q}\right) \label{Deltaeta}\\
V_0&=\frac{2\sqrt{G}}{c^2}\left[\frac{mc^2}{e}\sqrt{g_{00}}+\phi_q+\frac12\phi_e\right] \label{Deltaeps}
\end{align}
where we have used \eqref{defrs}-\eqref{EBIeletrostaticpot} and \eqref{Eptest}. Dropping a neutral body will surely not produce a naked singularity, since it increases the mass but not the black hole charge. We can see this in \eqref{Deltaeps} and \eqref{Deltaeta} noting that in this case $\Delta \eta=0$ and $\Delta \varepsilon<0$, which makes the system move vertically downward in the parameter space of Figure~\ref{fig:phasediag}. 

In terms of the phase diagram, violating the weak cosmic censorship conjecture means crossing the extremality curve given by $\varepsilon_{ex}(\eta)$ in~\eqref{eq:rex1}. From \eqref{Deltaeta}, we see that the absorption of the test body makes $\eta$ always decrease. Then, the condition can be rephrased as follows. If the parameter $\varepsilon$ decreases slower than $\varepsilon_{ex}(\eta)$, the system violates the weak cosmic censorship conjecture. Straightforward calculation gives
\begin{align}
\frac{\dd \varepsilon_{ex}}{\dd \eta}&=\frac{\varepsilon_{ex}}{2\eta}\left(\frac{f_{ex}-(1-\eta^2/4)}{f_{ex}+(1-\eta^2/4)}\right)
\label{dedeta}
\end{align}
This derivative is a monotonic function that starts at zero. We need to compare it to
\begin{align}
\frac{\Delta \varepsilon}{\Delta \eta}&= 
\frac{\varepsilon}{\eta}\left(\varepsilon V_0 -1 \right)\left(\frac{ 1+e/q}{1+e/q \ \varepsilon V_0}\right) \ .
\label{DeDeta}
\end{align}
The difficult term to evaluate is $V_0$ since it depends on the microphysics of the test body since it relates the mass $m$ and the charge $e$, and the nonlinearity of the potential $\phi_e$. However, we have good control of $\phi_q$ for all values of $\varepsilon$ and $\eta$. Let us define
\begin{equation}
U_q \equiv \frac{2\sqrt{G}}{c^2}\varepsilon\phi_q
=2\varepsilon^2\frac{ r_s}{r}{}_2F_1\left[\frac12,\frac14,\frac54,-\left(\frac{\varepsilon\sqrt{\eta}\ r_s}{r}\right)^4\right] \ .
\end{equation}
We shall now explore two possible physical scenarios for testing the WCCC: overcharging an extremal and a near-extremal EBI-BH.
\subsection{Extremal EBI-BH}

In the case of an extremal black hole,
a test body located just outside the extremal horizon has
\begin{align}\label{Uqex}
U_q& \approx
\frac{3 f_{ex}+(1-\eta^2/4)}{2f_{ex}+2(1-\eta^2/4)}
\quad \Rightarrow\quad 1\leq U_q \leq 1.5
\end{align}

From \eqref{Deltaeps}, we see that $\varepsilon V_0>U_q$, which implies that $\Delta\varepsilon<0$. Since $\Delta\eta<0$ we have $\frac{\Delta \varepsilon}{\Delta \eta}>0$. Furthermore, \eqref{DeDeta} is a monotonically increasing function of $\varepsilon V_0$. Therefore, since $\varepsilon V_0>U_q\geq 1$, 
\begin{equation}\label{DeDetaapp}
\frac{\Delta\varepsilon}{\Delta\eta}>
\frac{\varepsilon}{\eta}\left( U_q -1 \right)\left(\frac{ 1+e/q}{1+e/q \ U_q}\right)\approx \frac{\varepsilon}{\eta}\left( U_q -1 \right) \quad ,
\end{equation}
where the last equality comes from \eqref{Uqex} and $e/q\ll1$. The absorption of a charged body by the black hole implies $\Delta \eta<0$, so the system always moves from right to left in the parameter space. Thus, if $\Delta \varepsilon/\Delta \eta < \dd \varepsilon_{ex}/\dd \eta$, the system crosses the extremality line and violates the weak cosmic censorship conjecture. For an extremal EBI-BH, substituting \eqref{Uqex} into \eqref{dedeta} gives the exact relation
\begin{align}\label{dedetaU0}
\frac{\dd \varepsilon_{ex}}{\dd \eta}
=\frac{\varepsilon_{ex}}{\eta}\left(U_0-1\right) \quad ,
\end{align}
where
\begin{align}
U_0(\eta)=
\frac{3 f_{ex}+(1-\eta^2/4)}{2f_{ex}+2(1-\eta^2/4)} \quad .
\end{align}
Therefore, the preceding argument can be understood as a tangent criterion evaluated on the extremal curve. Namely, a physical process starting from an extremal configuration preserves the WCCC whenever its trajectory in parameter space has a slope not smaller than the slope of the extremal curve. Combining \eqref{DeDetaapp} with \eqref{dedetaU0}, at leading order in $e/q$, gives
\begin{align}\label{pwccc}
\frac{\Delta \varepsilon}{\Delta \eta}
>\frac{\varepsilon}{\eta}\left(U_q-1\right)
\rightarrow \frac{\varepsilon_{ex}}{\eta}\left(U_0-1\right)
=\frac{\dd \varepsilon_{ex}}{\dd \eta} \quad .
\end{align}
This result clarifies the scope of the extremal analysis: it compares the direction of the physical displacement with the tangent to the extremal curve. For a genuinely near-extremal initial black hole, however, the appropriate comparison is no longer the tangent at the initial value of $\eta$, but the secant line connecting the initial configuration to the extremal curve at the final value of $\eta$. We now turn to this case.

\subsection{Near-extremal EBI-BH}

Let us now analyze the case where the initial black hole lies inside the RN-phase, such that it is near extremality. In the phase diagram, the initial configuration is represented by the point $P_i=\paren{\eta,\varepsilon}$, such that $\varepsilon=\varepsilon_{ex}\paren{\eta}-d$ and $d>0$ is a small vertical distance on this phase space. 
After the charged particle is absorbed, the corresponding black-hole configuration is in the point $P_f=\paren{\eta_f,\varepsilon_f}=\paren{\eta+\Delta\eta,\varepsilon+\Delta\varepsilon}$ with $\Delta\eta<0$ given by Eq. \eqref{Deltaeta}.

Thus, in the $(\eta,\varepsilon)$ phase space, the physical trajectory moves toward smaller values of $\eta$. For a fixed displacement $\Delta\eta$, the limiting trajectory that saturates the WCCC is the one whose endpoint lies exactly on the extremal curve,
\begin{align}
P_{crit}=
\left(\eta+\Delta\eta,\varepsilon_{ex}(\eta+\Delta\eta)\right) \quad .
\end{align}
The line joining $P_i$ to $P_{crit}$ therefore defines the critical slope
\begin{align}\label{ScritDef}
S_{crit}
=
\frac{\varepsilon_{ex}(\eta+\Delta\eta)-\varepsilon}
{\Delta\eta}
=
\frac{\varepsilon_{ex}(\eta+\Delta\eta)-\varepsilon_{ex}(\eta)+d}
{\Delta\eta} \quad .
\end{align}
On the other hand, the actual absorption process has slope
\begin{align}
S_p\equiv\frac{\Delta\varepsilon}{\Delta\eta} \quad .
\end{align}
The near-extremal WCCC condition is consequently
\begin{align}
S_p\geq S_{crit} \quad .
\end{align}
If $S_p<S_{crit}$, the final state crosses the extremal curve and enters the naked-singularity region. If $S_{crit}\leq S_p<\dd\varepsilon_{ex}/\dd\eta$, the WCCC is still preserved, even though the tangent criterion alone would suggest otherwise. Finally, if $S_p\geq \dd\varepsilon_{ex}/\dd\eta$, the original tangent criterion is also satisfied.

To describe the corresponding horizon displacement, we introduce the dimensionless radial coordinate $\rho=r/r_s$, for which the time-time component of the metric is
\begin{align}
g(\rho,\varepsilon,\eta)=1-\frac{1}{\rho}
+\frac{\varepsilon^2}{\rho^2}
f\left(\frac{\rho}{\varepsilon\sqrt{\eta}}\right) \quad ,
\end{align}
The extremal horizon is then located at
\begin{align}
\rho_{ex}=\varepsilon_{ex}\sqrt{1-\frac{\eta^2}{4}} \quad .
\end{align}
Now we shall consider a near-extremal black hole and perturb around it. First, we have that $\varepsilon_{ex}(\eta)$ is the maximum allowed value of $\varepsilon$ for a fixed $\eta$, hence, a near-extremal black hole has 
\[
\varepsilon=\varepsilon_{ex}\paren{1-\delta_\varepsilon}\ , \quad \delta_\varepsilon\ll1
\]
On the other hand, we need to perturb the metric around the near-extremal horizon radius. Hence we expand $g\paren{\rho_+,\varepsilon,\eta}$ around the extremal horizon. Considering $\Delta\rho=\rho_+-\rho_{ex}$, we have
\[
g\paren{\rho_+,\varepsilon,\eta}=g\paren{\rho_{ex},\varepsilon_{ex},\eta}-\varepsilon_{ex} g_\varepsilon\vert_{ex}\delta_\varepsilon+g_\rho\vert_{ex}\Delta\rho+\frac{1}{2}g_{\rho\rho}\vert_{ex}\paren{\Delta\rho}^2+\ldots
\]
Since we are evaluating at the horizons $\rho_+$ and $\rho_{ex}$, we have $g(\rho_+)=g(\rho_{ex})=g_\rho\paren{\rho_{ex}}=0$. Hence
\be
\varepsilon_{ex} g_\varepsilon\vert_{ex}\delta_\varepsilon=\frac{1}{2}g_{\rho\rho}\vert_{ex}\paren{\Delta\rho}^2+\mathcal{O}\paren{\delta_\varepsilon^{3/2}}
\en
Therefore
\[
\rho_+=\rho_{ex}+\sqrt{2\varepsilon_{ex}\left.\frac{g_\varepsilon}{g_{\rho\rho}}\right|_{ex}}\sqrt{\delta_\varepsilon}=\rho_{ex}+h(\eta)\sqrt{\delta_\varepsilon}\ .
\]
Straightforward calculation results in
\[
h(\eta)=\sqrt{\frac{3}{2}} \sqrt{\frac{1+\frac{\eta ^2}{4}}{1-\frac{\eta ^2}{4}+2 \, _2F_1\left[\frac{1}{4},\frac{1}{2};\frac{5}{4};-\paren{\frac{1}{\eta}-\frac{\eta}{4}}^{-2}\right]}}
\]
Thus, we have that both perturbation parameters for $\epsilon$ and $\rho$ cannot be treated independently. We can therefore introduce the natural parameter 
\be
\alpha\equiv\sqrt{\delta_\varepsilon}
\en
so that our perturbation equations now read
\[
\begin{aligned}
\varepsilon&=\varepsilon_{ex}\paren{1-\alpha^2}\ ,\\
\rho_+&=\rho_{ex}+h\paren{\eta}\alpha+\mathcal{O}\paren{\alpha^2}\ .
\end{aligned}
\]
This expression shows that the radial displacement from the extremal horizon is of order $\alpha$, although the departure of $\varepsilon$ from $\varepsilon_{ex}$ is of order $\alpha^2$. Hence, the finite displacement considered for the starting point $P_i=\paren{\eta,\varepsilon_{ex}\paren{\eta}-d}$ is given by the perturbative parameter $\alpha$ as $d=\varepsilon_{ex}(\eta)\alpha^2$. Thus, the radial and charge-to-mass perturbations are tied to the same expansion parameter and should not be varied independently.

The finite size of the particle determines how close the body can be lowered before being released. If $\ell$ is its proper radius, the point of maximum approach satisfies
\begin{align}
\frac{\ell}{r_s}=\int_{\rho_+}^{\rho_a}
\frac{\dd \rho}{\sqrt{g(\rho,\varepsilon,\eta)}} \quad .
\end{align}
Near the horizon, $g(\rho,\varepsilon,\eta)\simeq g_\rho|_{\rho=\rho_+}(\rho-\rho_+)$, from which
\begin{align}
\rho_a=\rho_{ex}
+h(\eta)\left[
1+\frac14 g_{\rho\rho}\big|_{ex}
\left(\frac{\ell}{r_s}\right)^2
\right]\alpha+O(\alpha^2) \quad .
\end{align}
The coefficient appearing in this expression is
\begin{align}
\frac14 g_{\rho\rho}\big|_{ex}
=
\frac{2\left[
\eta^2/4
-2\,{}_2F_1\left(\frac14,\frac12;\frac54;-\frac{16\eta^2}{(\eta^2-4)^2}\right)
-1
\right]^2}
{9\left(1-\eta^4/16\right)} \quad .
\end{align}

We can now evaluate the black-hole electrostatic potential at this stopping point. Its expansion has the form
\begin{align}
U_q(\rho_a,\varepsilon,\eta)=U_0+U_1\alpha+O(\alpha^2) \quad ,
\end{align}
where
\begin{align}
U_1
=-\frac{2h(\eta)}{1-\eta^2/4}
\left[
1+\frac14 g_{\rho\rho}\big|_{ex}
\left(\frac{\ell}{r_s}\right)^2
\right]
\left[
{}_2F_1\left(\frac12,\frac14;\frac54;z_{ex}\right)
+\frac{2z_{ex}}{5}
{}_2F_1\left(\frac32,\frac54;\frac94;z_{ex}\right)
\right] \quad ,
\end{align}
with
\begin{align}
z_{ex}=-\frac{16\eta^2}{(4-\eta^2)^2} \quad .
\end{align}

Having fixed the geometric crossing criterion and the near-horizon expansion, we can now compare the two slopes. The expansion of $S_{crit}$ depends on how the particle charge scales with the distance from extremality. Writing $\lambda=e/q$, a balanced local limit is obtained by taking
\begin{align}
\lambda=\kappa\alpha \quad ,\qquad \kappa=O(1) \quad .
\end{align}
Then $\Delta\eta=-\eta\kappa\alpha+O(\alpha^2)$, and
\begin{align}\label{ScritSeries}
S_{crit}
=\frac{\dd \varepsilon_{ex}}{\dd\eta}
-\left[
\frac{\varepsilon_{ex}}{\eta\kappa}
+\frac{\eta\kappa}{2}
\frac{\dd^2\varepsilon_{ex}}{\dd\eta^2}
\right]\alpha
+O(\alpha^2) \quad .
\end{align}
On the physical-process side, using \eqref{DeDetaapp} together with $U_q=U_0+U_1\alpha+O(\alpha^2)$ gives
\begin{align}
S_p>
\frac{\dd \varepsilon_{ex}}{\dd\eta}
+\frac{\varepsilon_{ex}}{\eta}
\left[
U_1-\kappa(U_0-1)^2
\right]\alpha
+O(\alpha^2) \quad .
\end{align}
Combining this lower bound with \eqref{ScritSeries}, we obtain
\begin{align}\label{Ccriterion}
S_p-S_{crit}
>
\alpha\,\mathcal{C}(\eta,\kappa,\ell)
+O(\alpha^2) \quad ,
\end{align}
where
\begin{align}\label{Ccoef}
\mathcal{C}(\eta,\kappa,\ell)
=
\frac{\varepsilon_{ex}}{\eta}
\left[
U_1-\kappa(U_0-1)^2+\frac{1}{\kappa}
\right]
+\frac{\eta\kappa}{2}
\frac{\dd^2\varepsilon_{ex}}{\dd\eta^2} \quad .
\end{align}
Thus, for sufficiently small $\alpha$, the sign of $\mathcal{C}$ determines whether the physical lower bound on the process slope guarantees $S_p>S_{crit}$. We evaluated \eqref{Ccoef} numerically in the physical interval $0<\eta<2$, excluding the singular endpoint $\eta=2$. For $\ell/r_s=10^{-3}$, the minimum of $\mathcal{C}$ remains positive for $0<\kappa\lesssim 0.7$, whereas for larger values of $\kappa$ the coefficient becomes negative in part of the interval. The perturbative analysis therefore does not imply WCCC preservation for arbitrary particle charge. Rather, it identifies the quantitative domain
\begin{align}
\frac{e}{q}=\kappa\alpha \quad ,\qquad 0<\kappa\lesssim 0.7 \quad ,
\end{align}
for which the near-extremal EBI-BH remains on the black-hole side of the extremal curve at first nontrivial order. This result gives an explicit range of validity for the approximation: when the particle charge is too large compared with the distance from extremality, the first-order coefficient no longer guarantees WCCC preservation, and higher-order corrections or a breakdown of the test-body regime must be analyzed.

\bigskip

\section{Discussion} \label{sec:concl}
The Einstein-Born-Infeld black hole has four distinct physical behaviors. Each corresponds to a different region in the parameter space spanned by $\varepsilon=r_q/r_s$ and $\eta=r_\beta/r_q$. It can behave as a Schwarzschild black hole, as a Reissner-Nordstr\"om black hole, or as a naked singularity. An important feature is the possibility of overcharging the black hole for relatively large values of $\eta$. This makes the Einstein-Born-Infeld black hole an interesting physical system for testing two well-known conjectures, namely the Bekenstein entropy bound and the weak cosmic censorship conjecture. 

In order to test the latter, we investigated the absorption of a charged body by an extremal or near-extremal Einstein-Born-Infeld black hole. For the extremal case, we showed that the geometrical criterion is a tangent criterion on the extremal curve and that the corresponding slope can be written exactly as $\dd\varepsilon_{ex}/\dd\eta=(\varepsilon_{ex}/\eta)(U_0-1)$. We then extended the analysis to an initial configuration inside the RN-phase and near extremality. In this case, the tangent slope must be replaced by the critical secant slope connecting the initial black-hole configuration to the extremal curve at the final value of $\eta$. This construction fixes the perturbative ordering and gives a quantitative domain for the test-body approximation. Taking $e/q=\kappa\alpha$ and $\varepsilon=\varepsilon_{ex}(1-\alpha^2)$, the first nontrivial comparison is governed by the coefficient $\mathcal{C}(\eta,\kappa,\ell)$. Our numerical analysis shows that, for $\ell/r_s=10^{-3}$, this coefficient remains positive for $0<\kappa\lesssim0.7$, while it becomes negative for larger values of $\kappa$ in part of the interval $0<\eta<2$. Therefore, in the controlled perturbative domain $0<\kappa\lesssim0.7$, the near-extremal EBI-BH remains protected at first order in $\alpha$. Outside this domain, the first-order calculation no longer guarantees WCCC preservation, and one must include higher-order corrections or examine the limits of the test-body regime.

The results obtained in the present work are in accordance with previous results obtained using the Sorce-Wald method for this particular spacetime \cite{He2019}, and can be thought of as a phase-space geometric extension to testing the validity of WCCC. Recalling that the Sorce-Wald method is a rigorous, well established method relying on the validity of the Null Energy Condition (NEC) for the spacetime under inspection, our method can serve as an alternative to testing WCCC when spacetimes that do not fulfill the NEC possess a nontrivial phase structure. We intend to discuss such examples in a future publication.

\acknowledgments{
The authors would like to thank and acknowledge the financial support of the National Scientific and Technological Research Council (CNPq, Brazil), the State Scientific and Innovation Funding Agency of Esp\'\i rito Santo (FAPES, Brazil), and the Brazilian Federal Agency for Support and Evaluation of Graduate Education (CAPES, Brazil). MLP acknowledges support by the National Council for Scientific and Technological Development - CNPq and FAPERJ - Fundaçāo Carlos Chagas Filho de Amparo à Pesquisa do Estado do Rio de Janeiro, Processo SEI 260003/014960/2023.}

\end{document}